\documentclass[]{interact}

\usepackage{amsmath}
\usepackage{slashed}
\usepackage{txfonts}
\usepackage{microtype}
\usepackage{graphicx}
\usepackage[breaklinks=true]{hyperref}
\usepackage{color}
\usepackage{physics}
\usepackage{nicefrac}
\usepackage{soul}

\usepackage[numbers,sort&compress,merge]{natbib}
\bibpunct[, ]{[}{]}{,}{n}{,}{,}

\def\sout{\bgroup\markoverwith
{\textcolor{red}{\rule[0.5ex]{2pt}{0.5pt}}}\ULon}
\def\be{\begin{equation}}
\def\ee{\end{equation}}
\def\bes{\begin{equation*}}
\def\ees{\end{equation*}}
\def\bea{\begin{eqnarray}}
\def\eea{\end{eqnarray}}
\def\beas{\begin{eqnarray*}}
\def\eeas{\end{eqnarray*}}
\def\bal#1\eal{\begin{align}#1\end{align}}
\def\bals#1\eals{\begin{align*}#1\end{align*}}
\newcommand{\bk}[1]{\langle #1\rangle}

\renewcommand{\vec}{\vectorsym}
\newcommand{\del}{\partial}
\renewcommand*{\vec}[1]{\boldsymbol{#1}}

\graphicspath{{figures/}}

\begin{document}

 \title{Variational approaches to quantum impurities: from the Fr\"{o}hlich polaron to the angulon}


\author{
\name{Xiang Li, Giacomo Bighin, Enderalp Yakaboylu, Mikhail Lemeshko}
\affil{IST Austria (Institute of Science and Technology Austria), Am Campus 1, 3400 Klosterneuburg, Austria}
}

\maketitle

\begin{abstract}

Problems involving quantum impurities, in which one or a few particles are interacting with a macroscopic environment, represent a pervasive paradigm, spanning across atomic, molecular, and condensed-matter physics. In this paper we introduce new variational approaches to quantum impurities and apply them to the Fr\"{o}hlich polaron -- a quasiparticle formed out of an electron (or other point-like impurity) in a polar medium, and to the angulon -- a quasiparticle formed out of a rotating molecule in a bosonic bath.  We benchmark these approaches against established theories, evaluating their accuracy as a function of the impurity-bath coupling.

\end{abstract}

\begin{keywords}
Quantum impurity, quasiparticle, variational approach, molecular rotations, polaron, angulon.
\end{keywords}

\section{Introduction}

The concept of quasiparticle is one of the most fertile and far-reaching concepts in condensed-matter physics. When thinking in terms of quasiparticles one aims to describe collective excitations of a many-body system as effective emergent particles, hence the name~\cite{Venema:2016gd}.

One of the most well-known examples of quasiparticles is the Fr\"{o}hlich polaron, introduced by Landau~\cite{landau1933uber}, Pekar~\cite{pekar1946local}, and Fr\"{o}hlich~\cite{frohlich1954electrons} to describe the motion of electrons dressed by phonons in a polarisable medium. Over the years, the polaron became one of the standard, textbook models of condensed-matter physics, which has been studied using (and thereby spurred the development of) many theoretical approaches. Among those are  perturbative techniques~\cite{hubavc2010brillouin}, canonical transformations~\cite{LLP_53}, the Landau-Pekar strong-coupling approach~\cite{casteels2011strong}, Feynman's variational path integral method~\cite{feynman1955slow,Tempere_09}, as well as numerical techniques based on Monte Carlo~\cite{becker1983monte,Mishchenko_00} and renormalization group~\cite{Grusdt_2015}.

Notably, the polaron  concept  has proven useful far beyond the original physics problem (electrons in crystals), and was successfully applied  to  systems as diverse as electrons on the surface of liquid helium~\cite{devreese2007frohlich,jackson1981polaronic}, doped antiferromagnetic Mott insulators~\cite{dagotto1994correlated}, magnetic semiconductors~\cite{kaminski2002polaron}, and ultracold gases \cite{Tempere_09}. In the quasiparticle picture, the polaron accounts for the effect of the many-body environment on the quantum impurity by means of the renormalisation of the particle parameters -- such as its energy and mass. In such a way, the effect of $\sim10^{23}$ particles of the bath can be understood in terms of a handful of renormalised parameters -- a drastic simplification, which in many cases allows to obtain extremely accurate results.

All  quantum impurities described by the  Fr\"{o}hlich polaron model are structureless (such as electrons) or can be considered structureless (such as atoms whose electronic structure is not perturbed by their surroundings). A compelling question is whether molecules and -- in general -- more complex quantum systems can be described as quantum impurities using the quasiparticle approach. Recently, a new quasiparticle, the angulon, has been introduced to describe a molecule interacting with a bosonic many-body field, such as a superfluid~\cite{Lemeshko_2015,LemSchmidtChapter, lemeshko2016quasiparticle}. While angulons can be thought of as ``rotational analogues'' of polarons, there are several important differences. First, as opposed to translational motion, rotations in three-dimensional space are described by a non-Abelian $\mathrm{SO}(3)$ algebra, which leads to intricate theoretical machinery of angular momentum addition. Furthermore,  anisotropic molecular geometry results in anisotropic impurity-boson coupling, which renders many-body interactions explicitly dependent on the molecular orientation. The unique properties of such a system motivated the introduction of new analytical~\cite{LemSchmidtChapter, Bighin_2017,yakaboylu2018theory} and numerical techniques~\cite{Bighin:2018}, which can be applied  to the Fr\"{o}hlich polaron as well.

In this paper, inspired by the recent advances in polaron theory~\cite{Pietro_14,Devreese15,grusdt2015renormalization,Grusdt_2015} as well as by the recent developments concerning angulons~\cite{Lemeshko_2015, LemSchmidtChapter, PhysRevX.6.011012}, we introduce new variational methods for the Fr\"{o}hlich polaron and for the angulon. In particular, we introduce two variational approaches based on a single-phonon expansion either over the ground-state or after a canonical transformation,  leading to two different non-perturbative descriptions of the Fr\"{o}hlich polaron, as well as a diagonalization technique based on the well-known Pekar ansatz \cite{pekar1946local}, that we dub `Pekar diagonalization'.  The results we obtain are benchmarked against Feynman's all-coupling theory \cite{Feynman:1955zz} and against the Pekar ansatz \cite{pekar1946local}.

The paper is organized as follows. In Sec.~\ref{sec_hamiltonian}, we first briefly introduce the Fr\"{o}hlich Hamiltonian. Then, we introduce two new variational ansaetze for the polaron problem. Namely, in Sec.~\ref{sec_chevy} we study a variational ansatz based on a single-phonon excitation over the ground state, and in Sec.~\ref{sec_prx} we discuss a variational ansatz based on a single-phonon excitation on top of a bosonic coherent state,  in order to extend the description to the intermediate- and strong-coupling regimes. In Sec.~\ref{sec_pekar} we introduce a new diagonalisation method based on the Pekar ansatz  and apply it to the Fr\"{o}hlich polaron and to the angulon. The conclusions of the paper are drawn  in Sec.~\ref{sec_conc}.

\section{Fr\"{o}hlich Hamiltonian} \label{sec_hamiltonian}

The Fr\"{o}hlich Hamiltonian, describing an impurity immersed in a bosonic bath, is given by:
\be
\label{ham}
\hat H_F = \frac{\vec{\hat{P}}^2}{2m}  + \sum_{\vec{k}} \omega(k) \hat b^\dagger_{\vec{k}} \hat b_{\vec{k}}  + \sum_{\vec{k}} V(k) \left( e^{-i \vec{k}\cdot \vec{\hat{x}}}\hat b^\dagger_{\vec{k}} + e^{i \vec{k}\cdot \vec{\hat{x}}} \hat b_{\vec{k}} \right) \, .
\ee
Here the first term represents the kinetic energy of an impurity with mass $m$. The second term, with $\sum_{\vec{k}} \equiv \int d^3 k / (2 \pi)^3$, corresponds to the kinetic energy of the bosons, as parametrised by the dispersion relation $\omega(k)$. The bosonic creation and annihilation operators, $\hat b^\dagger_{\vec{k}}$ and $\hat b_{\vec{k}}$, obey the commutation relation $[\hat b_{\vec{k}}, \hat b^\dagger_{\vec{k}'}] = (2 \pi)^3 \delta (\vec{k} - \vec{k}')$. Finally, the last term is the impurity-bath interaction, where $V(k)$ determines  the coupling strength, and $\vec{\hat{x}}$ is the position operator of the impurity with respect to the laboratory frame.

In what follows, we use Fr\"{o}hlich's original parameters, i.e.\ a constant dispersion relation for gapped optical phonons, $\omega(k) = \omega_0$, and the coupling strength,
\be
\label{frohlich_coup}
V(k) = \sqrt{\frac{2^{3/2}\pi\alpha}{k^2}} \, ,
\ee 
$\alpha$ being the electron-phonon coupling constant in units of $m=\omega_0=\hbar \equiv 1$. The Hamiltonian of Eq. (\ref{ham}) possesses translational symmetry, which follows from the fact that the total linear momentum of the system, 
\be
\label{total_p}
\hat{\vec{\Pi}} = \hat{\vec{P}} + \sum_{\vec{k}} \vec{k} \hat b^\dagger_{\vec{k}} \hat b_{\vec{k}} \, ,
\ee
commutes with the Hamiltonian~(\ref{ham}). Conservation of the total linear momentum allows us to label the polaron quasiparticle with the momentum quantum number.

\section{Single phonon expansion} \label{sec_chevy}

\begin{figure}[h!]
  \centering
  \includegraphics[width=0.70\linewidth]{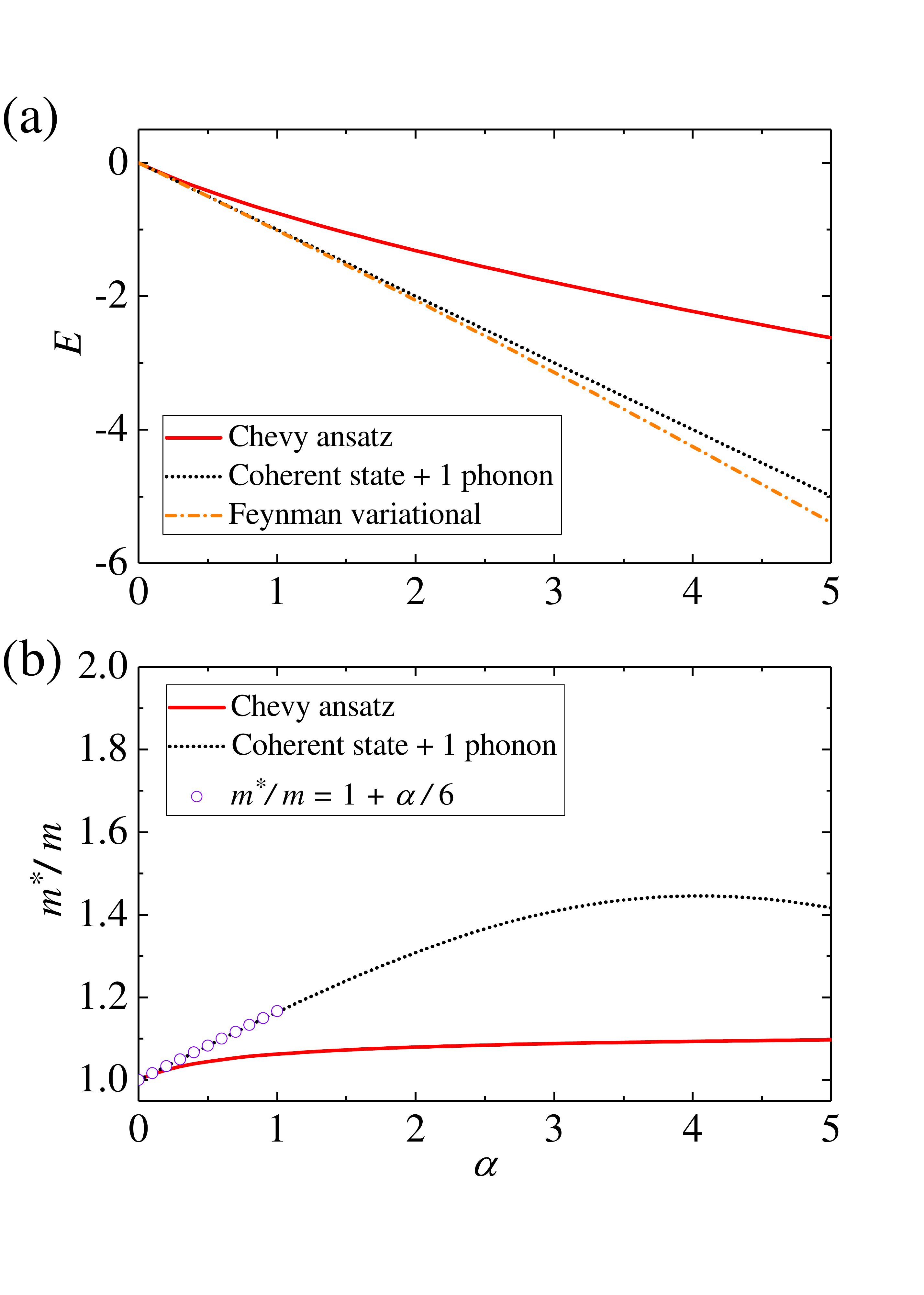}
 \caption{(a) The polaron energy as a function of the Fr\"{o}hlich coupling constant, $\alpha$, for the Chevy ansatz,~Eq.(\ref{eq:polaron_chevy}) (red solid line), coherent state on top of single phonon excitation,~Eqs.(\ref{eq:1phonon_wavefunction}) and~(\ref{eq:coherent_ansatz}) (black dotted line), and the Feynman variational method~\cite{Feynman:1955zz} (orange dash-dotted line). (b)  Renormalization of the polaron mass as a function of the Fr\"{o}hlich coupling constant, $\alpha$,  for the Chevy ansatz (red solid line), coherent state on top of single phonon excitation (black dot line),  and the weak coupling theory~\cite{Devreese15} (purple circles). See the text. }
 \label{fig_energy}
\end{figure}

Inspired by the so-called `Chevy ansatz', originally introduced for an imbalanced Fermi-gas~\cite{Chevy_06,ngampruetikorn2012repulsive,lan2014single}, we expand the state vector up a single phonon excitation. Taking into account the conservation of the total linear momentum, we write down the following variational ansatz:
\be
\label{eq:polaron_chevy}
\ket{\psi_{\vec{p}}} = \sqrt{Z_{\vec{p}}}\ket{\vec{p}} \ket{0} + \sum_{\vec{k}} \beta_{\vec{p}} (\vec{k}) \ket{\vec{p}-\vec{k}} \hat b^\dagger_{\vec{k}} \ket{0} \, ,
\ee
where $\sqrt{Z_{\vec{p}}} $ and $\beta_{\vec{p}} (\vec{k})$ are variational parameters with the normalization condition $\sqrt{Z_{\vec{p}'}}^* \sqrt{Z_{\vec{p}}} + \sum_{\vec{k}} \beta_{\vec{p}'} (\vec{k})^* \beta_{\vec{p}} (\vec{k}) = \delta(\vec{p}'-\vec{p})$. Minimization of the functional $\bra{\psi_{\vec{p}'}} \hat H_F - E \ket{\psi_{\vec{p}}} $ with respect to the parameters $\sqrt{Z_{\vec{p}}}^*$ and $\beta_{\vec{p}} (\vec{k})^*$ yields the following coupled equations
\bal
\label{var_z}
\frac{\del F}{\del \sqrt{Z_{\vec{p}}}^*} & = |Z_{\vec{p}}| \left(\frac{\vec{p}^2}{2m} - E \right) + \sum_{\vec{k}}\beta_{\vec{p}}(\vec{k}) V(k) = 0 \, ,\\
\label{var_b}
\frac{\del F}{\del \beta_{\vec{p}}(\vec{k})^*} & = \beta_{\vec{p}}(\vec{k}) \left(\frac{\left(\vec{p}-\vec{k}\right)^2}{2m} + \omega(k) - E \right) + \sqrt{Z_{\vec{p}}} V(k) = 0 \, .
\eal
If we substitute $\beta_{\vec{p}} (\vec{k})$ from Eq.~\eqref{var_b} into Eq.~\eqref{var_z}, we obtain the Dyson equation
\be
\label{energy_one_phonon}
E = \frac{\vec{p}^2}{2m} - \Sigma_{\vec{p}}(E) \, ,
\ee
which can be solved to obtain the variational energy $E$. The self-energy is given by
\be
\label{eq:self_energy}
\Sigma_{\vec{p}} (E) = \sum_{\vec{k}} \frac{V(k)^2}{(\vec{p}-\vec{k})^2/(2m) + \omega(k) - E} \, ,
\ee
which can be solved self-consistently. Combining the variational energy $E$, the normalization condition, Eq. (\ref{var_z}) and Eq. (\ref{var_b}), one can obtain the values of variational coefficients $\sqrt{Z_{\mathbf{p}}}$ and $\beta_{\mathbf{p}}(\mathbf{k})$. We note that the self-energy of Eq.(\ref{eq:self_energy}) coincides with that obtained by means of field-theoretical approaches in Ref. \cite{Grusdt_2015}. 

In the iterative solution to Eq.~\eqref{energy_one_phonon}, the leading-order term is given by $E^{(1)} = \vec{p}^2/(2M)$, and the second-order term reads $E^{(2)}= \vec{p}^2/(2M)  - \Sigma_{\vec{p}}(E^{(1)})$, which matches the result of second order perturbation theory. Therefore, the variational energy~(\ref{energy_one_phonon}) is non-perturbative as it corresponds to resummation over all diagrams describing single-phonon excitations, see Refs.~\cite{LemSchmidtChapter, Bighin_2017} for further details.

 Fig.~\ref{fig_energy} (a) shows the Fr\"ohlich polaron energy as calculated from Eq.~(\ref{energy_one_phonon}). A comparison with Feynman's all-coupling theory \cite{Feynman:1955zz} shows that, despite the inherently non-perturbative nature of a Chevy-like ansatz, in the case of the Fr\"ohlich polaron its effectiveness in determining the ground state energy is limited to the weak-coupling region. In addition to this, Fig.~\ref{fig_energy} (b), present results for the renormalized polaron mass $m^*$, defined by
\be
\frac{1}{m^*} = \left. \frac{\del^2 E}{\del \vec{p}^2} \right|_{\vec{p}=0} \, .
\ee
 Here, except for very small values of the coupling $\alpha$, our Chevy-like ansatz deviates from the classical perturbation-theory result $m^*/m = 1 + \alpha/6$, tending to a constant value for sufficiently large $\alpha$. The scope of applicability of the present treatment, however, in the light of the results for the energy presented in Fig.~\ref{fig_energy} (a), should not be extended to that region.

In this section, we have shown that the variational ansatz of Eq. (\ref{eq:polaron_chevy}) yields a good prediction of ground energy in weak coupling region through a simple, fully analytical calculation. Moreover, working with a variational ansatz makes the underlying physics clear: the variational coefficient $\sqrt{Z_{\vec{p}}}$ is the quasiparticle weight, i.e. a measure of the overlap between the dressed impurity and a bare particle, whereas the variational coefficient $\beta_{\mathbf{p}}(\mathbf{k})$ contains information about the occupation of phonon states.

\section{Coherent state on top of single phonon excitation} \label{sec_prx}

 Recently a new variational ansatz has been introduced in order to tackle the angulon problem \cite{PhysRevX.6.011012} in the limit of a slowly-rotating impurity.  This method is based on single phonon excitation expansion after a coherent state transformation that brings the Hamiltonian to a diagonal form in the limit of a  slowly rotating impurity. Aiming to use this method for the Fr\"ohlich polaron, we start by applying the Lee-Low-Pines transformation~\cite{LLP_53},
\be
\hat T = \exp\left( - i \vec{\hat{x}} \cdot \sum_{\vec{k}} \vec{k} \hat b^\dagger_{\vec{k}} \hat b_{\vec{k}} \right) \, ,
\ee
after which the Fr\"{o}hlich Hamiltonian then can be written as
\be
\hat H_F'  = \hat{T}^{-1} \hat H_F \hat{T} = \frac{1}{2m} \left(\vec{\hat{P}} -  \sum_{\vec{k}} \vec{k} \hat b^\dagger_{\vec{k}} \hat b_{\vec{k}}\right)^2  + \sum_{\vec{k}} \omega(k) \hat b^\dagger_{\vec{k}} \hat b_{\vec{k}}  + \sum_{\vec{k}} V(k) \left( \hat b^\dagger_{\vec{k}} + \hat b_{\vec{k}} \right) \, ,
\ee
commuting with $\vec{\hat{P}}$, i.e., $[\hat H_F',\vec{\hat{P}}]=0$. Then, the corresponding state vector can be written as a product state,
\be
\label{eq:1phonon_wavefunction}
\ket{\Phi_{\vec{p}}} = \ket{\varphi} \otimes \ket{\vec{p}} \, .
\ee
a similar approach having been introduced in Ref. \cite{shchadilova2016quantum}. Here the state vector $\ket{\vec{p}}$, with $\vec{p}$ being the total momentum number of the impurity-bath system in the laboratory-frame, corresponds to the impurity wave function, while the bosonic state $\ket{\varphi}$ refers to the bosonic part of the following Hamiltonian
\be
\label{bosonic_ham}
\hat H_F' = \frac{\vec{p}^2}{2m}  + \sum_{\vec{k}} \tilde{\omega}(\vec{k}) \hat b^\dagger_{\vec{k}} \hat b_{\vec{k}}  + \sum_{\vec{k}} V(k) \left( \hat b^\dagger_{\vec{k}} + \hat b_{\vec{k}} \right) + \frac{1}{2m} \hat{\Gamma} \, ,
\ee
where $\tilde{\omega}(\vec{k}) = \omega(k) - \vec{k} \cdot \vec{p}/m +k^2/(2m) $, and $\hat{\Gamma} = \sum_{\vec{k}, \vec{k}'} \vec{k} \cdot \vec{k}'  \hat b^\dagger_{\vec{k}}  \hat b^\dagger_{\vec{k}'} \hat b_{\vec{k}} \hat b_{\vec{k}'} $. In the limit of $m \to \infty$, the Hamiltonian~\eqref{bosonic_ham} can be diagonalized using the following coherent state transformation
\be
\hat U = \exp\left(- \sum_{\vec{k}} \frac{V(k)}{\tilde{\omega}(\vec{k})} ( \hat b^\dagger_{\vec{k}} - \hat b_{\vec{k}} ) \right) \, .
\ee
After applying this transformation to Eq.~\eqref{bosonic_ham} we obtain
\be
\hat{H}_F{''} = \hat U^{-1} \hat H_F'  \hat U =  \frac{\vec{p}^2}{2m}  + \sum_{\vec{k}} \tilde{\omega}(\vec{k}) \hat b^\dagger_{\vec{k}} \hat b_{\vec{k}}  -  \sum_{\vec{k}} \frac{V(k)^2}{\tilde{\omega}(\vec{k})} + \frac{1}{2m} \hat U^{-1} \hat{\Gamma}  \hat U \, .
\ee

Next we introduce the following variational ansatz for the bosonic state:
\be
\label{eq:coherent_ansatz}
\ket{\varphi} = g \ket{0} + \sum_{\vec{k}} \alpha(\vec{k})  \hat b^\dagger_{\vec{k}}  \ket{0} \, .
\ee
Then, minimization of the functional $F = \bra{\varphi} \hat{H}_F{''} - E \ket{\varphi} $ with respect to the parameters $g^*$ and $\alpha(\vec{k})^*$ gives the following system of equations
\bal
\label{var1}
\frac{\del F}{\del g^*} & = -g \tilde{E} - \frac{1}{m} \sum_{\vec{k}, \vec{k}'}\alpha(\vec{k}) \left( \frac{V(k')}{\tilde{\omega}(\vec{k}')} \right)^2 \frac{V(k)}{\tilde{\omega}(\vec{k})} \vec{k} \cdot \vec{k}' = 0 \\
\nonumber \frac{\del F}{\del \alpha(\vec{k})^*} & =- \frac{g}{m} \sum_{\vec{k}'} \left( \frac{V(k')}{\tilde{\omega}(\vec{k}')} \right)^2 \frac{V(k)}{\tilde{\omega}(\vec{k})} \vec{k} \cdot \vec{k}' +  \alpha(\vec{k}) \left[- \tilde{E} +  \tilde{\omega}(\vec{k}) + \frac{1}{m} \sum_{\vec{k}'} \vec{k} \cdot \vec{k}'  \left( \frac{V(k')}{\tilde{\omega}(\vec{k}')} \right)^2 \right] \\
\label{var2}
& + \frac{1}{m}\sum_{\vec{k}'} \alpha(\vec{k}') \vec{k} \cdot \vec{k}' \frac{V(k')}{\tilde{\omega}(\vec{k}')} \frac{V(k)}{\tilde{\omega}(\vec{k})} = 0 \,,
\eal
where
\be
\tilde{E} = E - \frac{\vec{p}^2}{2m} + \sum_{\vec{k}} \frac{V(k)^2}{\tilde{\omega}(\vec{k})} - \frac{1}{2m} \sum_{\vec{k}, \vec{k}'} \vec{k} \cdot \vec{k}' \left( \frac{V(k)}{\tilde{\omega}(\vec{k})} \right)^2 \left( \frac{V(k')}{\tilde{\omega}(\vec{k}')} \right)^2 \, .
\ee
We further use the rotational symmetry of the problem, and, without loss of generality, assume that $\mathbf{p} \parallel \hat{z}$. Then, solving $\alpha(\vec{k})$ from Eq.~\eqref{var2} as function of $g$ and plugging into Eq.~\eqref{var1} gives us Dyson equation
\be
\label{eq:polaron_1phonon}
E = \frac{\vec{p}^2}{2m} - \Sigma_{\vec{p}} (E) \, ,
\ee
from which one can solve for the variational energy $E$. The self-energy here has the following form
\be
\Sigma_{\vec{p}} (E) = \sum_{\vec{k}} \frac{V(k)^2}{\tilde{\omega}(\vec{k})} - \frac{1}{2m} I_z^2 +  A_z I_z \,.
\ee
Moreover, we have defined
\be
I_z=\sum_\mathbf{k} k_z \left(V(k)/\tilde{\omega}(\vec{k}) \right)^2
\ee
and
\be
A_z =  \frac{I_z}{m} \sum_\mathbf{k} \frac{k_z^2}{m} \frac{(V(k)/\tilde{\omega} (\mathbf{k}))^2}{- \tilde{E} +  \tilde{\omega}(\vec{k}) + k_z I_z/m} \left(1+\sum_\mathbf{k} \frac{k_z^2}{m} \frac{(V(k)/\tilde{\omega} (\mathbf{k}))}{- \tilde{E} +  \tilde{\omega}(\vec{k}) + k_z I_z/m} \right)^{-1} \, .
\ee
Of course, in the limit of $m\to \infty$, we obtain the deformation energy of the bath, $ E = - \sum_{\vec{k}} V(k)^2/\omega(k)$.

 In Fig.~\ref{fig_energy} (a), we study the resulting polaron energy as a function of the Fr\"{o}hlich coupling constant, $\alpha$. The treatment developed in the present Section provides an energy estimate remarkably better than the Chevy-like ansatz introduced in the previous Section, and in particular the energy is considerably close to Feynman's all-coupling theory \cite{Feynman:1955zz} over a broad range of values of $\alpha$. In addition to this, Fig.~\ref{fig_energy} (b) shows the renormalization of the polaron mass as a function of $\alpha$, the result of the approach developed in the present Section being considerably larger than that obtained in previous Section, coinciding with the the perturbation-theory result $m^*/m = 1 + \alpha/6$ up to $\alpha \sim 1$.

\section{Pekar Diagonalization} \label{sec_pekar}
\subsection{Polaron} \label{subsec_polaron}

\begin{figure}[h!]
  \centering
  \includegraphics[width=0.70\linewidth]{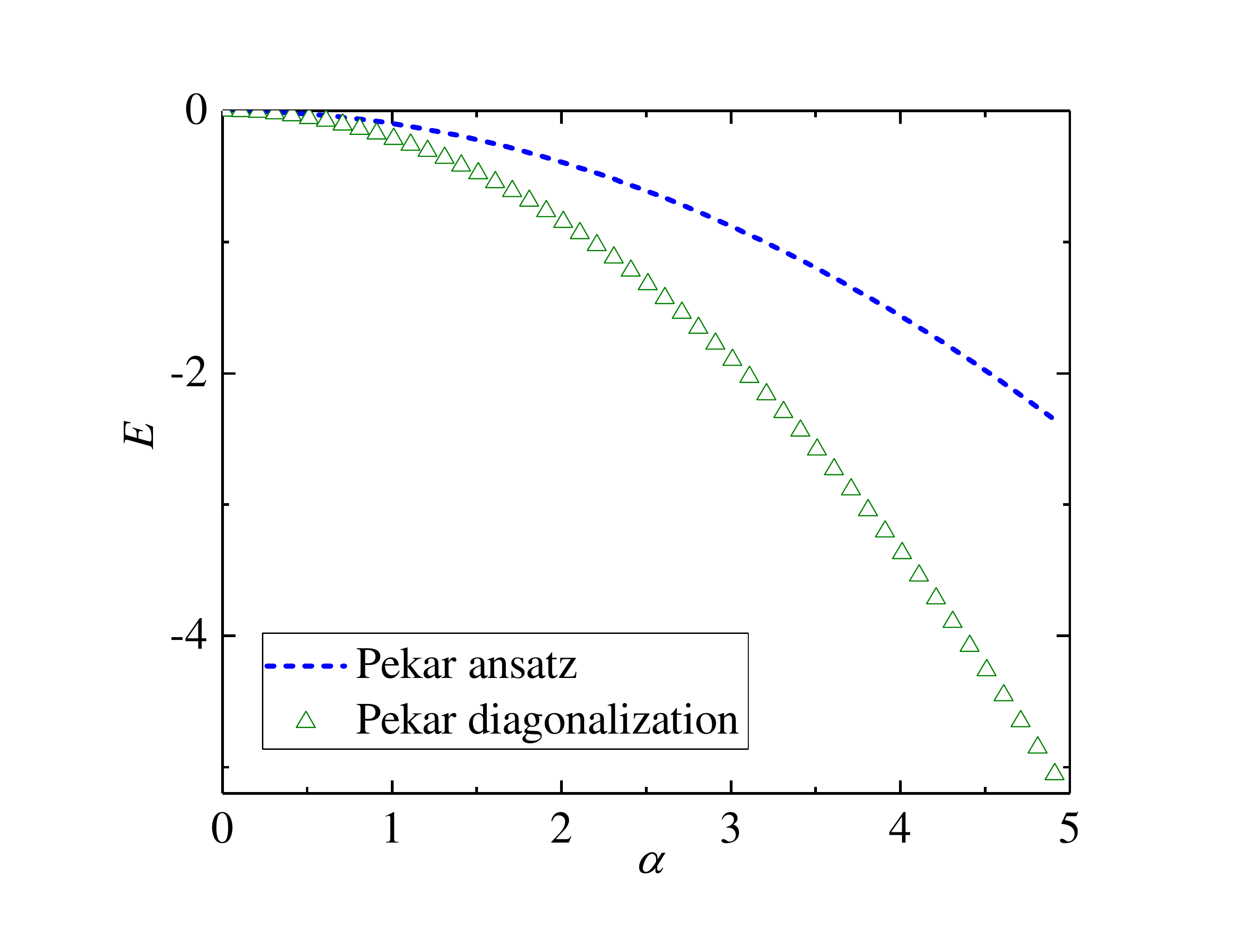}
 \caption{The polaron energy as a function of the Fr\"{o}hlich coupling constant, $\alpha$, for the Pekar ansatz,~Eq.(\ref{eq:pekar_ansatz}) (blue dash line), and the Pekar diagonalization technique,~Eqs.(\ref{eq:pekar_diag_state_vector}) and~(\ref{eq:pekar_diag_trial}) (green triangles). See the text.}
 \label{fig_pekar}
\end{figure}

The strong-coupling theory of the Fr\"{o}hlich Hamiltonian can be studied within the Pekar ansatz~\cite{pekar1946local, Devreese15}:
\be
\label{eq:pekar_ansatz}
\ket{\Psi_P} = \ket{\varphi} \otimes \ket{\xi_B} \,,
\ee
where $\ket{\varphi}$ and $\ket{\xi_B}$  correspond to the impurity wavefunction and the bosonic state, respectively.  The Pekar treatment that we are now going to briefly review   essentially corresponds to the Born-Oppenheimer approximation. It is assumed that the phonons and the impurity have two completely different timescales, or, more precisely, that the phonons can adjust instantaneously as the slowly moving impurity changes its position. In order to carry out this plan one takes the expectation value, $\bra{\varphi} \hat H_F \ket{\varphi}$, the resulting effective bosonic Hamiltonian can be diagonalized using the following coherent-state transformation:
\be
\hat U = \exp\left[-\sum_{\vec{k}} \frac{V(\vec{k})}{\omega(k)}\left(  \bk{ e^{-i \vec{k}\cdot \hat{\vec{x}}}} \hat b^\dagger_{\vec{k}}- \text{H.c.}\right) \right] \, ,
\label{eq:Upolaron}
\ee
where $\bk{\hat A} \equiv \bra{\varphi} \hat A \ket{\varphi}$. The bosonic state minimizing the Pekar energy is given by $ \ket{\xi_B} = \hat U \ket{0} $, and the respective ground-state energy is:
\be
\label{pekar_energy}
\varepsilon_0 = \frac{1}{2m} \bk{\vec{\hat{P}}^2}  - \sum_{\vec{k}} \frac{|V(\vec{k}) \bk{ e^{-i \vec{k}\cdot \hat{\vec{x}}}}|^2}{\omega(k)} \, .
\ee 
In general, the impurity wavefunction for the ground state can be modeled by the following radial Gaussian function~\cite{Devreese15}
\be
\varphi (\vec{x}) =  \left( \frac{\beta}{\pi} \right)^{3/4}e^{-\beta r^2/2} \, .
\ee
Minimization of the Pekar energy~\eqref{pekar_energy} with respect to the variational parameter $\beta$ yields~\cite{Devreese15}
\be
\varepsilon_0 = - \frac{\alpha^2}{3\pi} \, .
\ee

In what follows  we present an extension of the Pekar approach that we dub `Pekar diagonalization'. For this purpose, we introduce the following state vectors
\be
\label{eq:pekar_diag_state_vector}
\ket{\Psi_n} = \ket{\varphi_n} \exp({-\hat X_{nn}}) \ket{0} \, ,
\ee
where 
\be
\hat X_{nn} = \sum_{\vec{k}} \frac{ V(\vec{k})}{\omega(k)}\left( \bk{ e^{-i \vec{k}\cdot \hat{\vec{x}}}}_{nn} \hat b^\dagger_{\vec{k}}- \text{H.c.}\right) \, ,
\ee
with $\bk{\hat A}_{nm} \equiv \bra{\varphi_n} \hat A \ket{\varphi_m}$ and $\bk{\hat A}_n \equiv \bra{\varphi_n} \hat A \ket{\varphi_n}$.
Then, the corresponding matrix element can be written as
\bal
\label{pekar_diag}
& H_{F\, nm} \equiv \bra{\Psi_n} \hat H_F  \ket{\Psi_m} = \frac{e^{-\Gamma_{nm}}}{2m} \bk{\vec{\hat{P}}^2}_{nm}  + e^{-\Gamma_{nm}} \sum_{\vec{k}} \frac{V(\vec{k})^2}{\omega(k)}\times \\
\nonumber &  \left( N_{nm} \bk{ e^{i \vec{k}\cdot \hat{\vec{x}}}}_{nn} \bk{ e^{-i \vec{k}\cdot \hat{\vec{x}}}}_{mm} - \bk{e^{-i \vec{k}\cdot \hat{\vec{x}}}}_{nm} \bk{e^{i \vec{k}\cdot \hat{\vec{x}}}}_{nn} - \bk{e^{i \vec{k}\cdot \hat{\vec{x}}}}_{nm} \bk{e^{-i \vec{k}\cdot \hat{\vec{x}}}}_{mm} \right) \, ,
\eal
where we define $N_{nm} \equiv \braket{\varphi_n}{\varphi_m}$, and
\be
\Gamma_{nm} = \frac{1}{2}\sum_{\vec{k}} \left(\frac{ V(\vec{k})}{\omega(k)}\right)^2 \left(\bk{ e^{-i \vec{k}\cdot \hat{\vec{x}}}}_{nn} \bk{ e^{i \vec{k}\cdot \hat{\vec{x}}}}_{mm}-\bk{ e^{i \vec{k}\cdot \hat{\vec{x}}}}_{nn} \bk{ e^{-i \vec{k}\cdot \hat{\vec{x}}}}_{mm} \right) \,
\ee
Naturally, the diagonal terms correspond to Eq.~\eqref{pekar_energy}. We note that a similar diagonal technique has been applied in ultracold fermionic and bosonic mixtures \cite{cao2017unified,mistakidis2018repulsive}.

In order to use the diagonalization technique~\eqref{pekar_diag}, we use the following ansatz for the impurity wave function~\cite{hagen1970excited}
\be
\label{eq:pekar_diag_trial}
\varphi_n (\vec{x}) = N_n e^{-\beta r} (1 + a_1 r + \cdots a_n r^n) \, ,
\ee
corresponding to $s$-wave states. Here $\beta$ and $a_n$ are the variational parameters with $n$ labeling excited states. After finding the optimum values of the variation parameters for each excited state, we can diagonalize Eq. \eqref{pekar_diag}. In Fig.~\ref{fig_pekar}, we show the corresponding energy, where we  use only 2 basis vectors. It can be seen that the Pekar diagonalization technique remarkably improves the Pekar ansatz~Eq.(\ref{eq:pekar_ansatz}) in the strong-coupling region, and more rigorous results can be given with larger matrix or with a better trial state $\varphi_n(x)$.

\subsection{Angulon} \label{subsec_angulon}

\begin{figure}[h!]
  \centering
  \includegraphics[width=0.70\linewidth]{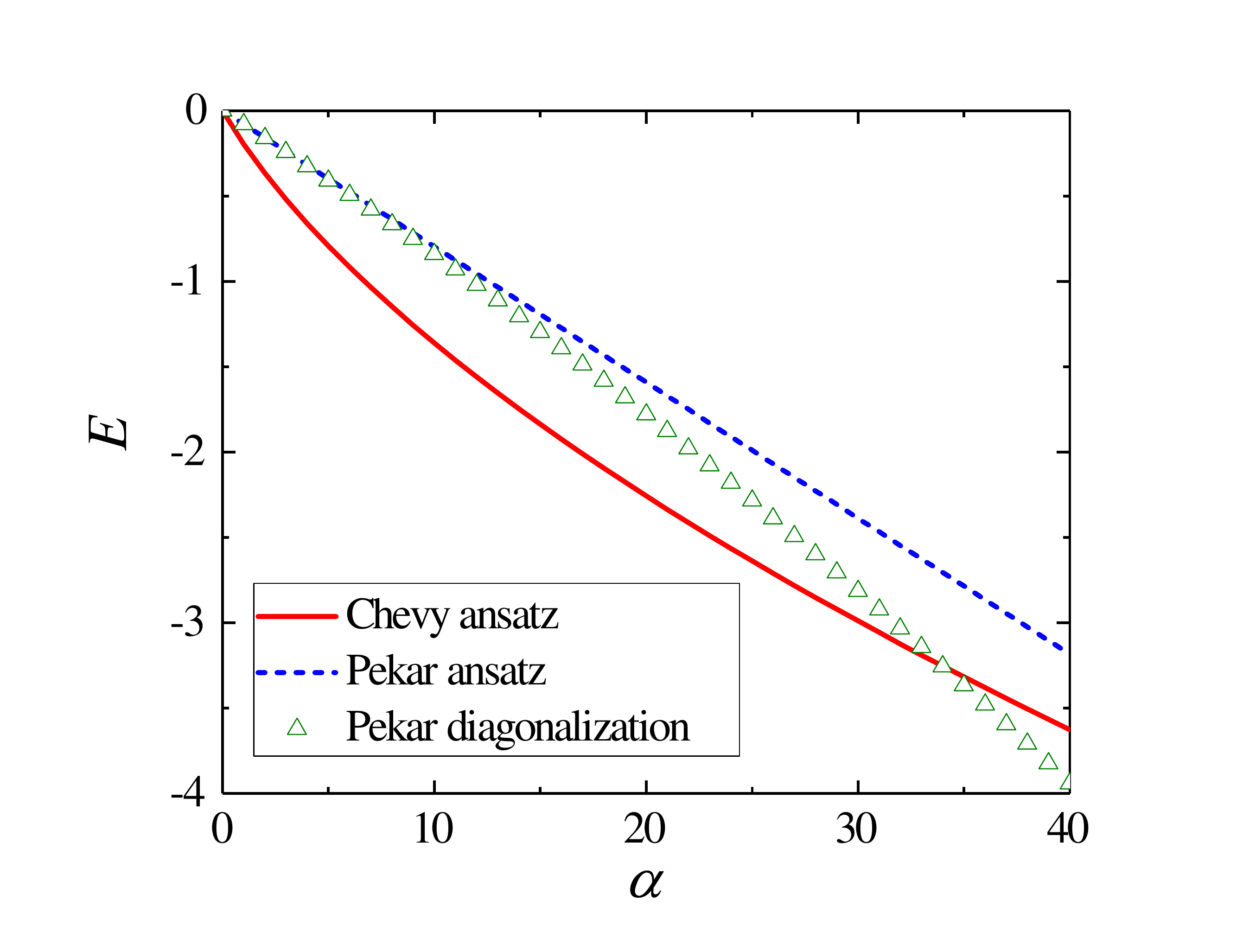}
 \caption{The angulon ground state energy as a function of the angulon coupling constant, $\alpha$, for the Chevy ansatz~\cite{Lemeshko_2015, LemSchmidtChapter} (red solid line), the Pekar ansatz \cite{Li_2017} (blue dashed line), and the Pekar diagonalization method of Eq. (\ref{eq:ansatz_angulon}) (green triangles). The basis consists of the vectors with $j=0,1,2$. See the text.}
 \label{fig:angulon}
\end{figure}

As a next step, we show that the Pekar diagonalization we have just introduced can be applied to the angulon, i.e. a quasiparticle describing a quantum molecular impurity with rotational degrees of freedom. In order to do so, let us introduce the angulon Hamiltonian \cite{Lemeshko_2015,LemSchmidtChapter}:
\begin{eqnarray}
   \hat{H_A}&=&B\hat{\mathbf{J}}^2 + \sum_{k\lambda\mu}\omega(k)\hat{b}^\dag_{k\lambda\mu}\hat{b}_{k\lambda\mu} +  \sum_{k\lambda\mu}U_\lambda(k)\left[Y^*_{\lambda\mu}(\hat{\Omega})\hat{b}^\dag_{k\lambda\mu}+ \text{H.c.} \right] 
   \label{eq:Hangulon}
\end{eqnarray}
describing a molecular impurity -- schematised as a rigid rotor exchanging angular momentum with a bosonic many-body environment. Let us briefly discuss the structure of Eq. (\ref{eq:Hangulon}). In the first term, expressing the rotational kinetic energy of the molecular impurity, $B$ and $\hat{\mathbf{J}}$ are the rotational constant and the angular momentum operator, respectively. The second term of Eq. (\ref{eq:Hangulon}) represents the kinetic energy of the non-interacting bosons with dispersion relation $\omega(k)$; the bosonic creation and annihilation operators, $\hat{b}^\dag_\mathbf{k}$ and $\hat{b}_\mathbf{k}$, are expressed in the angular momentum basis: $\hat{b}^\dag_{k\lambda\mu} = k(2\pi)^{-3/2}\int d\Omega_k\hat{b}^\dag_k i^\lambda Y^*_{\lambda\mu}(\Omega_k)$, while $\lambda$ and $\mu$ define the boson angular mementum and its projection onto the laboratory-frame $z$ axis, see Ref. \cite{LemSchmidtChapter} for more details. Finally, the third term of Eq.(\ref{eq:Hangulon}) describes the impurity-bath interaction, where the coupling potential, $U_\lambda(k)$, parametrises the interaction of impurity with bosons carrying angular momentum $\lambda$ and linear momentum $k$.

To apply the Pekar diagonalization technique to the angulon, we consider the following basis vector
\begin{equation}
    |\Psi_{jm}\rangle = |jm \rangle\exp[-\hat{X}_{jm}]|0\rangle ,
    \label{eq:ansatz_angulon}
\end{equation}
where the free rotor eigenstates, $|jm\rangle$, are labeled by the angular momentum, $j$, and its projection, $m$, on the laboratory $z$ axis. Furthermore, in writing Eq. (\ref{eq:ansatz_angulon}) we introduced $ \hat{X}_{jm} $ defined as follows
\begin{equation}
    \hat{X}_{jm} = \sum_{k\lambda\mu}\frac{U_\lambda(k)}{\omega(k)}\left[\langle j m \vert Y^*_{\lambda\mu}(\hat{\Omega}) \vert j m \rangle~\hat{b}^\dag_{k\lambda\mu} - \text{H.c.}\right].
\end{equation}
Following the scheme outlined in Section \ref{sec_pekar}, we obtain for the angulon
\begin{eqnarray}
    \langle j'm'|H_A|jm\rangle &=& Be^{-\Gamma_{j'm',jm}}\langle j'm'| \hat{\mathbf{J}}^2|jm\rangle + e^{-\Gamma_{j'm',jm}}\sum_{k\lambda\mu}\frac{U^2_\lambda(k)}{\omega(k)}\times
    \nonumber\\
    & &\Big[\langle j'm'|Y_{\lambda\mu}(\hat{\Omega})|j'm'\rangle\langle  jm|Y^*_{\lambda\mu}(\hat{\Omega})|jm\rangle \delta_{j'j}\delta_{m'm}
    \nonumber\\
    & &- \langle j'm'|Y^*_{\lambda\mu}(\hat{\Omega})|jm\rangle\langle j'm'|Y_{\lambda\mu}(\hat{\Omega})|j'm'\rangle
    \nonumber\\
    & &- \langle j'm'|Y_{\lambda\mu}(\hat{\Omega})|jm\rangle\langle jm|Y^*_{\lambda\mu}(\hat{\Omega})|jm\rangle\Big]
    \label{eq:matrix_element}
\end{eqnarray}
where
\begin{eqnarray}
    \Gamma_{j'm',jm} &=& \frac{1}{2}\sum_{k\lambda\mu}\left(\frac{U_\lambda(k)}{\omega(k)}\right)^2\Big(\langle j'm'|Y^*_{\lambda\mu}(\hat{\Omega})|j'm'\rangle\langle jm|Y_{\lambda\mu}(\hat{\Omega})|jm\rangle
    \nonumber\\
    & &- \langle j'm'|Y_{\lambda\mu}(\hat{\Omega})|j'm'\rangle\langle jm|Y^*_{\lambda\mu}(\hat{\Omega})|jm\rangle\Big)
\end{eqnarray}
becomes zero due to the symmetry of Clebsch-Gordan coefficients \cite{Varshalovich}. It is worth noting that this is due to the basis vector we chose, see Eq. (\ref{eq:ansatz_angulon}), and would not necessarily be zero for other choices of basis vectors.

As a simplifying assumption, here we ignore the detailed structure of the anisotropic interaction potential, introducing the following dimensionless impurity-bath interaction parameters:
\begin{equation}
    \alpha_\lambda = \sum_k\frac{U^2_\lambda(k)}{\omega_kB}.
\end{equation}
 and assuming $U_\lambda(k)\equiv U(k)$, and therefore $\alpha_{\lambda}\equiv\alpha$.

In Fig. \ref{fig:angulon} we compare the results of the Pekar diagonalization technique with the `standard' Pekar approach \cite{pekar1946local,Li_2017} and with the Chevy ansatz for the angulon \cite{Lemeshko_2015, LemSchmidtChapter}. One can see that, over the whole range of couplings we consider, the Pekar diagonalization technique leads to a lower variational ground state-energy than the standard Pekar approach, which only considers the diagonal term of Hamiltonian, i.e taking only $j'=j$ and $m'=m$ in Eq. (\ref{eq:matrix_element}). Fig. \ref{fig:angulon} also shows that, beyond a critical coupling strength the technique gives a lower ground state energy with respect to Chevy ansatz \cite{Lemeshko_2015,Cherepanov} .

The Pekar diagonalization technique, as compared with the `standard' Pekar approach, is particularly powerful in the angulon case as a consequence of the non-Abelian $\mathrm{SO}(3)$ algebra describing the coupling of angular momenta. More precisely: a phonon coupling two impurity states with angular momentum $j$ and $j'$ will have an angular momentum $\lambda$ in the range $\{~|j'-j|,|j'-j|+1,...,j'+j-1,j'+j~\}$, thereby leading to a number of nonzero off-diagonal terms in Eq. (\ref{eq:matrix_element}).  The technique we have introduced allows one to obtain more accurate estimates since it accounts for these off-diagonal entries, as opposed to the `standard' Pekar treatment. This is particularly evident when higher angular momenta are considered; in Fig. \ref{fig:angulon}, $j=0,1,2$ and $\lambda=0,1,2,3,4$ due to the selection rules imposed by the Clebsch-Gordan coefficients.

\section{Conclusions}  \label{sec_conc}

 In this paper we  introduced analytic approaches to quantum impurity problems, namely two variational ansaetze and a new diagonalization approach that we called `Pekar diagonalization'. The results of the variational techniques were compared with well-established benchmarks such as the Pekar ansatz -- as far as the strong-coupling regime is concerned -- and Feynman's all-coupling variational theory. As expected, an approach inspired by the Chevy ansatz works accurately for smaller values of the coupling whereas approaches based on the Pekar ansatz are reliable in the strong-coupling region. On the other hand, the approximation involving a single-phonon excitation on top of a coherent state transformation provides an estimate remarkably close to Feynman's all-coupling theory in a wide parameter region.  A promising future direction consists in using such an ansatz for other polaron problems beyond the Fr\"ohlich model, as well as for other quantum impurity problems.
 
 
 We have also exemplified the Pekar diagonalization technique by studying the ground energy of both the polaron and the angulon quasiparticles. The results have shown that the diagonalization technique we developed here represents a an improvement compared to the `standard' Pekar ansatz over a wide range of coupling strengths, especially in  the strong-coupling region. Pekar diagonalization represents a promising approach to quantum impurities, especially for systems -- such as the angulon -- where the `standard' Pekar approach can not provide reliable results.
    
\section{Acknowledgements}
We are grateful to Areg Ghazaryan for valuable discussions. This work was supported by the  Austrian Science Fund (FWF), Project No. P29902-N27. E. Y. acknowledges financial support received from the People Programme (Marie Curie Actions) of the European Union's Seventh Framework Programme (FP7/2007-2013) under REA grant agreement No. [291734].
 
\bibliography{ap.bib}
\bibliographystyle{tfo}

\end{document}